\documentclass[final,3p,times]{elsarticle}
\usepackage{amssymb}
\usepackage{epstopdf}
\usepackage{mathrsfs}
\usepackage{stmaryrd}
\biboptions{sort&compress}%,longnamesfirst,angle,semicolon}
\usepackage{graphicx,amsfonts,listings,xcolor,colortbl}
\usepackage{multirow}
\usepackage{booktabs}
\usepackage{amsmath}
\usepackage{colortbl}
\usepackage{color}
\usepackage{rotating}
\usepackage{subfigure}
\usepackage[normalem]{ulem}
\usepackage{algorithm}
\usepackage{algorithmic}

\newtheorem {definition} {\bf Definition}[section]

\usepackage{dcolumn}% Align table columns on decimal point
\usepackage{bm}% bold math
\usepackage{array}
\begin{document}

\begin{frontmatter}



\title{Identifying super-spreaders in information-epidemic coevolving dynamics on multiplex networks}%\tnoteref{NSFC}}
%\tnotetext[NSFC]{This work is supported by the National Science Foundation of China
%(No.~60873108).}


\author[addr1]{Qi Zeng\fnref{equ}}
\author[addr1]{Ying Liu\corref{cor1}\fnref{equ}}
\ead{shinningliu@163.com}
\cortext[cor1]{Corresponding author}
\author[addr2,addr3]{Ming Tang}
%\author[addr2,addr4]{Ming Tang\corref{cor1}}
%\ead{tangminghan007@gmail.com}
\fntext[equ]{These authors contributed equally to this paper.}
%\cortext[cor2]{Corresponding author}
\author[addr1]{Jie Gong}
%\author[addr1]{Yang Wang}

\address[addr1]{School of Computer Science, Southwest Petroleum University, Chengdu 610500, P. R. China}
\address[addr2]{School of Physics and Electronic Science, East China Normal University, Shanghai 200241, China}
\address[addr3]{Shanghai Key Laboratory of Multidimensional Information Processing, East China Normal University, Shanghai 200241, China}

\begin{abstract}
Identifying super-spreaders in epidemics is important to suppress the spreading of disease especially when the medical resource is limited. In the modern society, the information on epidemics transmits swiftly through various communication channels which contributes much to the suppression of epidemics. Here we study on the identification of super-spreaders in the information-disease coupled spreading dynamics. Firstly, we find that the centralities in physical contact layer are no longer effective to identify super-spreaders in epidemics, which is due to the suppression effects from the information spreading. Then by considering the structural and dynamical couplings between the communication layer and physical contact layer, we propose a centrality measure called coupling-sensitive centrality to identify super-spreaders in disease spreading. Simulation results on synthesized and real-world multiplex networks show that the proposed measure is not only much more accurate than centralities on the single network, but also outperforms two typical multilayer centralities in identifying super-spreaders. These findings imply that considering the structural and dynamical couplings between layers is very necessary in identifying the key roles in the coupled multilayer systems.
\end{abstract}

\begin{keyword}
multiplex network\sep information-disease coupled spreading dynamics\sep super-spreader\sep coupling-sensitive centrality

\PACS 89.75.Fb \sep 87.23.Ge \sep 89.75.-k
%\pacs{89.75.-k}{Complex systems}
%\pacs{05.10.-a}{Computational methods in statistical physics and nonlinear dynamics}
%\pacs{89.75.Fb}{Structures and organization in complex system}
%\pacs{05.45.-a}{Nonlinear dynamics and chaos}
%\pacs{87.19.lo}{Information theory}
%\pacs{87.23.Ge}{Dynamics of social systems}
%\pacs{89.65.-s}{Social and economic systems}
%\pacs{87.19.X-}{Diseases}
%\pacs{89.75.Hc}{Networks and genealogical trees}

\end{keyword}

\end{frontmatter}

\section{Introduction}
Identification of the most important nodes in complex networks is an active field in network science~\cite{borgatti2006, pei2018}. The important nodes in different contexts have their specific implications, such as the opinion leaders that can influence the public views in the social media~\cite{watts2007}, the critical neurons and regions in brain networks associated with brain functions~\cite{del2018}, the most influential spreaders that can maximize information diffusion or epidemic spreading~\cite{kempe2003,kitsak2010,morone2015}, and the articulation nodes that maintain the integrity and connectivity of networks~\cite{braunstein2016,tian2017,morone2019}. The most commonly used method to identify important nodes in networks is the use of centrality measures which evaluate nodes' importance from the local or global structure of the networks, such as degree~\cite{freeman1978}, eigenvector centrality~\cite{bonacich2001}, closeness centrality~\cite{sabi1966}, betweenness centrality~\cite{freeman1977}, PageRank~\cite{brin1998}, k-shell index obtained from k-core decomposition~\cite{bolobas1984} and the nonbacktracking centrality~\cite{martin2014}. The centrality-based methods are very successful and find their wide applications in identifying kinds of important nodes~\cite{lu2016,jalili2017}.

In spreading dynamics, the most important nodes called influential spreaders or super-spreaders are nodes which can induce the largest outbreak sizes when the spreading originates from them. As control of epidemics is a major challenge the human beings are facing with~\cite{anderson1992,zhang2020}, identification of influential spreaders is a key step in optimizing the available resources and ensuring more efficient control strategy~\cite{kitsak2010}. A great many of methods are proposed to identify influential spreaders~\cite{radicchi2016,liu2017,min2018,deng2020,jalili2020}. However these progresses are mostly in single layer networks, while some real complex systems are better represented as multilayer networks~\cite{mikko2014}. For example an individual may have relationships with others in different ways, either being friends, colleagues, schoolmates, or doing business, which is a multilayer network~\cite{szell2010}. Integrating these different types of interactions into a single network may lose some critical information of the system. The multilayer network approach is adopted in understanding the robustness of infrastructures~\cite{buldyrev2010,gao2012}, coevolving spreading dynamics of information and epidemics~\cite{clara2013}, evolutionary games~\cite{perc2013}, functions of brain~\cite{morone2017} and stability of economical and financial systems~\cite{pole2015}.

When an epidemic outbreaks in the contact population, information on the epidemic is easily transmitted through kinds of channels, such as the online social platform and mobile communication network. The information on epidemic disease promotes people to adopt self-protection measures to reduce their risks of being infected, thus helping to suppress the diffusion of epidemic. Meanwhile, the wide spread of epidemic further enhances the spreading of information~\cite{funk2009}. These can be considered as two processes on a two-layer multiplex network where the information diffuses on the communication layer and the disease spreads on the physical contact layer, and nodes are the individuals~\cite{clara2013}. It is found that the asymmetrical interplay between layers has significant impact on the spreading dynamics on top of the multiplex networks, such as changing the epidemic threshold and suppressing the infected population in the stationary state~\cite{manlio2016,arruda2018,wang2016,yang2019}. As for searching for the influential spreaders, it is natural to consider that the spreading influence of a node in the physical contact layer will be reduced due to the information spreading on the communication layer, but the extent of reduction depends on the structure and dynamics on the communication layer.

To address the issue of identifying the most important nodes in multiplex networks, a lot of centralities and methods are proposed, such as the multiplex PageRank~\cite{halu2013,iaco2016}, multiplex eigenvector centrality~\cite{sola2016,taylor2021}, multiplex betweenness~\cite{albert2014}, tensor-decomposition based methods~\cite{wang2017}, and methods based on the local or global structure of the networks~\cite{azimi2014,domenico2015,basaras2019}. These methods provide more accurate rankings than approaches based on the aggregated networks or single-layer networks. However, these methods mainly focus on the multilayer structure, while neglecting the dynamical couplings between layers. Researches have pointed out that the influence of a node in the spreading process is a result of the interplay between dynamics and network structure~\cite{klemm2012,radicchi2016-2}, and dynamic-sensitive measures are more efficient in identifying super-spreaders than structural measures in single networks and interconnected networks~\cite{zhao2014,liu2016}.

In this paper, we work on identifying the most influential spreaders in the coupled information-epidemic dynamics. By considering three dynamical and structural couplings between two layers, which are the two-layer relative spreading rate, the inter-layer coupling strength and the inter-layer degree correlation, we propose a measure called coupling-sensitive centrality (CS) to identify super-spreaders in the coevolving dynamics on multiplex networks. Simulation results show that the CS centrality is not only much more accurate than the centralities of degree, eigenvector centrality, k-shell index and PageRank in the contact layer, but is also more accurate than two typical multiplex centralities, which are the multiplex PageRank and multiplex eigenvector centrality. While being applied to a variety of real-world multiplex networks, the proposed measure also works well, implying that considering the couplings between layers is crucial in finding key nodes in multilayer systems.

The rest of the paper is organized as follows. In section 2, we give the preliminaries. In section 3, we propose the coupling-sensitive centrality. In section 4, we represent the effectiveness of the proposed centrality in synthesized networks and real-world networks. Finally we give a conclusion in section 5.
\section{Preliminaries}
In this part, we first describe the information-epidemic coupled spreading model on a two-layer multiplex network. Then we give a brief description of the benchmark centralities of degree, eigenvector centrality, k-shell index and PageRank on single layer network, and the multiplex PageRank and multiplex eigenvector centrality on multiplex network, which are used as competitors to our measure to identify super-spreaders. Finally we give the definition of the imprecision function and Kendall's tau correlation coefficient, which are two methods to evaluate the performance of different measures in identifying super-spreaders.
\subsection{The coupled information-epidemic model on multiplex network}
To describe the information-epidemic coevolving dynamics, we use the SIR-SIRV spreading model in ref.~\cite{wang2016}. In this model, a multiplex network is composed of two layers where the nodes represent individuals and links represent their interactions. The information spreads on the upper communication layer A and the disease spreads on the bottom physical contact layer B. In the communication layer, the classical susceptible-infected-recovered (SIR) model is used to describe the information spreading process. In the SIR model, a node can be in one of the three states: susceptible (S), infected or informed in information spreading (I) and recovered (R). At each time step, an I-state node infects or informs all its susceptible neighbors with rate $\beta_A$ and then recovers with rate $\mu_A$. The spreading stops until there is no I-state node in the network. In the physical contact layer where disease spreads, the same SIR dynamics are adopted with an additional state vaccinated (V), where nodes of V-state will neither be infected nor transmit disease. The disease transmission rate is $\beta_B$ and the recovering rate of infected nodes is $\mu_B$.  The effective transmission rates for layer A and B are respectively $\lambda_A=\beta_A/\mu_A$ and $\lambda_B=\beta_B/\mu_B$.

\begin{figure}
\begin{center}
\includegraphics[width=13.5cm]{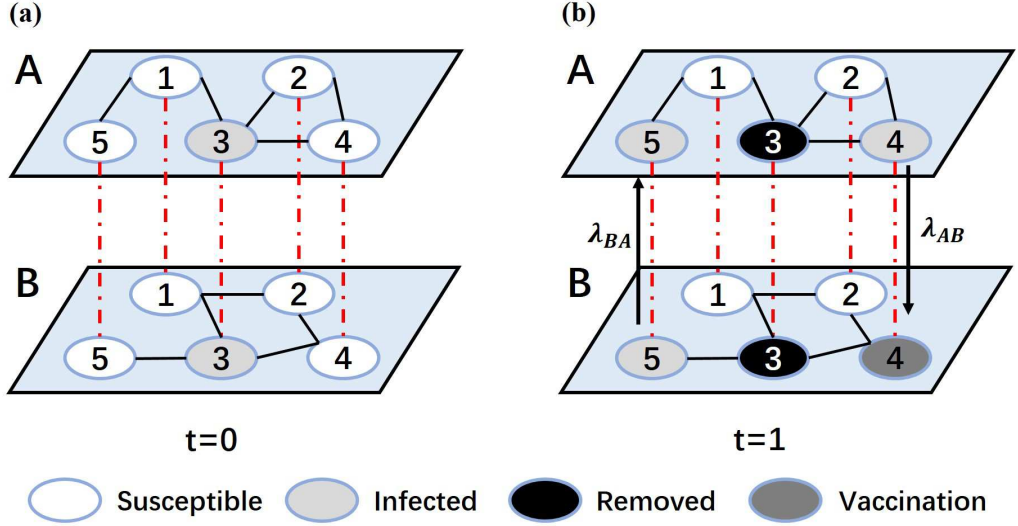}
\caption{The schematic representation of the information-epidemic coupled dynamics on a multiplex network. Layer A is the communication network where information spreads, and layer B is the physical contact network where disease spreads. (a) At $t=0$, a node is informed and infected and all other nodes are in susceptible states in both layers. The informed and infected node transmits information and disease to its neighbors in layer A and B respectively with rate $\lambda_A$ and $\lambda_B$. (b) The informed node in layer A makes its counterpart in layer B be vaccinated with rate $\lambda_{AB}$, and the infected node in layer B makes its counterpart in layer A be informed with rate $\lambda_{BA}$.}
\label{figure1}
\end{center}
\end{figure}

The dynamical coupling of the two layers is as follows. For an informed node $i_A$ in layer A, if its counterpart $i_B$ in layer B is in S-state, then $i_B$ transfers to state V with immunization rate $\lambda_{AB}$. This immunization rate represents the willingness or capability of the informed individuals to get vaccinated. The larger $\lambda_{AB}$ is, the more counterparts of the informed nodes will get vaccinated and the suppression effect from information layer is more significant. For an infected node $j_B$ in layer B, if its counterpart $j_A$ in layer A is in S-state, then $j_A$ transfers to state I with rate $\lambda_{BA}$. The $\lambda_{BA}$ represents the probability that an infected individual is aware of the epidemic or is willing to transmit information on it. The larger $\lambda_{BA}$ is, the more nodes in layer A become aware of the epidemic and the information spreads more widely. A schematic representation of the SIR-SIRV model is shown in Fig.~\ref{figure1}.

In simulations, each layer of the multiplex network is a scale-free network with power law degree distribution $p(k)\thicksim k^{-\gamma}$. We generate layer A from the uncorrelated configuration model (UCM) with the number of nodes $N=10000$, the power exponent $\gamma=2.6$ and the average degree $<k>=6$. The minimal degree is set as $k_{min}=3$ and the maximal degree is set as $k_{max}=\sqrt{N}$. Then we generate layer B by keeping the same node sets of layer A. The degree sequence of layer B is generated by copying the degree sequence of layer A first and then exchanging the degrees of randomly selected pairs of nodes until the specified inter-layer degree correlation is reached. The multiplex network is thus a set of nodes with two different types of connections represented  by two layers. The inter-layer degree correlation of the two layers is quantified by the Spearman rank correlation coefficient~\cite{lee2012}.
\begin{definition}\rm
Inter-layer degree correlation. The inter-layer degree correlation is defined as
\begin{equation}
m_s=1-6\frac{\sum_{i=1}^N \Delta_i^2}{N(N^2-1)},
\end{equation}
where $N$ is the number of nodes, $\Delta_i$ is the rank difference of node $i$ in two ordered sequences ranked by degree in layer A and layer B respectively. $m_s$ ranges in [-1,1]. If the degrees of nodes in two layers are positively correlated, $m_s \to 1$. If the degrees of nodes in two layers are negatively correlated, $m_s\to -1$. If the degrees of nodes in two layers are uncorrelated, $m_s \to 0$. The degree correlation of a node in two layers has a potential impact on the spreading influence of the node in the contact layer. From the centrality's perspective, a node with larger degree has a greater spreading influence in the network. For nodes with the same degree in the contact layer B, their degrees in the communication layer A imply different suppression effects of information on epidemic spreading, making the centrality in the contact layer be unable to predict the spreading influence of nodes in disease spreading.
\end{definition}

In the coevolving dynamics, initially a seed node is infected on layer B and its counterpart in layer A is informed. The information and disease spread in layer A and layer B respectively until a final state is reached. As we focus on disease spreading, we take the proportion $R_i$ of recovered nodes in the contact layer in the steady state as the spreading influence of the seed node $i$. The disease transmission rate $\lambda_B$ is set to be $0.13$ for the synthesized network which is three times of the epidemic threshold of layer B. As indicated in ref.~\cite{liu2015}, the transmission rate will not impact the relative ranking of node influence if it is above the epidemic threshold and within a few times of epidemic threshold. The recovery rates are set as $\mu_A=\mu_B=1$ for simplicity. The relative spreading rate of layer A and layer B is $\gamma_{AB}^{\lambda}=\lambda_A/\lambda_B$, which reflects the dynamical coupling of the two layers. If $\gamma_{AB}^{\lambda}=\lambda_A/\lambda_B$ is large enough, we can consider that the information spreads first and the disease spreads on the residual network after vaccination. All results in simulations are obtained by averaging over 100 runs.
\subsection{Competitor centralities}
\subsubsection{Centralities on single-layer network}
Consider a network $G(V,E)$, where $V$ is the set of nodes, $E$ is the set of edges, and $N=|V|$ is the number of nodes. The adjacent matrix of $G$ is $A_{N*N}=a_{ij}$, where $a_{ij}=1$ if there is an edge between node $i$ and $j$, otherwise $a_{ij}=0$. The degree $k$ quantifies how many direct neighbors a node has, which is $k_{i}=\sum_{j\in V\backslash i}a_{ij}$. Degree is the simplest centrality measure in quantifying node importance.
\begin{definition}\rm
Eigenvector centrality. The eigenvector centrality of node $i$ is defined as
\begin{equation}\label{eigenvector}
\ e_i=\lambda^{-1}\sum_{n=1}^Na_{ij}e_j,
\end{equation}
which gives $\lambda e=Ae$ in matrix notation. Here $e$ is the right leading eigenvector corresponding to the largest eigenvalue $\lambda$ of the adjacent matrix A. The eigenvector centrality takes into account both the quantity and quality of neighbors in determining the importance of a node.
\end{definition}

\begin{definition}\rm
K-shell index. The k-shell index $ks$ is obtained in the k-shell decomposition process. Initially, nodes with degree $k=1$ are removed from the network together with their links. After removing all nodes with $k=1$, some nodes initially with degree more than one may have only one link left. Continue to remove them until no node with degree one is left. The removed nodes are assigned a k-shell value $ks=1$. Then nodes with degree $k\leq2$ are removed in the same way and given a value $ks=2$. The pruning process continues until all nodes are removed and given a $ks$ value. Nodes with large $ks$ are considered to be in the core of the network and are super-spreaders.
\end{definition}

\begin{definition}\rm
PageRank Centrality. The PageRank centrality is famous for its success in the web ranking technology used by Google corporation. It is defined as
\begin{equation}\label{pagerank}
PR_i(t)=(1-c)\sum_{j=1}^Na_{ij}\frac{PR_j(t-1)}{k_j^{out}}+\frac{c}{N},
\end{equation}
where $k_j^{out}$ is the out-degree of a neighbor node $j$, and $c$ is a free parameter to represent the random jump of users not visiting along the links. The algorithm iterates until a steady-state and the $PR_i$ for each node is obtained.
\end{definition}
\subsubsection{Centralities on multiplex network}
First we introduce the Functional Multiplex PageRank (FMPR) defined in ref.~\cite{iaco2016}, which is a generation of PageRank centrality to the multiplex networks. This centrality considers the weights of different types of connections between two nodes. Suppose a multiplex network $\overrightarrow{G}=(G_1, G_2,...,G_M)$ is composed of a set $V$ of $N$ nodes and $M$ layers. Each layer is $G_\alpha=(V, E_\alpha)$, where $E_\alpha$ is the set of edges in layer $\alpha$ and $\alpha=1,2,...,M$. The adjacent matrix element $a_{ij}^{\alpha}=1$ if node $i$ and $j$ has a link in layer $\alpha$, otherwise $a_{ij}^{\alpha}=0$. To characterize the multiplex network with overlap, a vector called multilink $\overrightarrow{m}^{ij}$ is defined, where $\overrightarrow{m}^{ij}=(m_1, m_2,...,m_M)$ has elements $m_{\alpha}=0,1$. The number of possible types of multilink is $2^M$. A pair of nodes $i$ and $j$ are connected by a multilink $\overrightarrow{m}^{ij}$ if and only if $\overrightarrow{m}^{ij}=(a_{ij}^{[1]}, a_{ij}^{[2]},..., a_{ij}^{[M]})$. Then the multiadjacency matrices $\textbf{A}^{\overrightarrow{m}}$ defines the connection of nodes with multilink $\overrightarrow{m}^{ij}$, with elements $A_{ij}^{\overrightarrow{m}}=1$ if the node pair $i$ and $j$ is connected by a multilink $\overrightarrow{m}^{ij}$, otherwise $A_{ij}^{\overrightarrow{m}}=0$. Therefore $A_{ij}^{\overrightarrow{m}}$ can be expressed as
\begin{equation}\label{Am}
A_{ij}^{\overrightarrow{m}}=\prod_{\alpha=1}^M[m_{\alpha}a_{ij}^{[\alpha]}+(1-m_{\alpha})(1-a_{ij}^{[\alpha]})].
\end{equation}
For the two-layer multiplex network the multi-adjacent matrices are
\begin{equation}\label{Am2}
\begin{aligned}
&A_{ij}^{(1,0)}=a_{ij}^{[1]}(1-a_{ij}^{[2]})\\
&A_{ij}^{(0,1)}=(1-a_{ij}^{[1]})a_{ij}^{[2]}\\
&A_{ij}^{(1,1)}=(a_{ij}^{[1]}a_{ij}^{[2]}).
\end{aligned}
\end{equation}
\begin{definition}\rm
The Functional Multiplex PageRank centrality $X_i(\textbf{z})$ of node $i$ is defined as
\begin{equation}\label{Xz}
X_i(\textbf{z})=\widetilde{\alpha}\sum_{j=1}^NA_{ij}^{\overrightarrow{m}^{ij}}z^{\overrightarrow{m}^{ij}}\frac{1}{k_j}X_j+\beta v_i,
\end{equation}
where $\widetilde{\alpha}$ is the possibility that in the steady state a random walker jumps from node $j$ to a neighbor $i$, otherwise it jumps to a randomly connected node in the multiplex network. $\textbf{z}$ is a tensor with elements $z^{\overrightarrow{m}}\geq0$ associated to every type of multilink $\overrightarrow{m}$ representing its influence, and $z^{\overrightarrow{0}}=0$. A random walker jumps to a neighbor along the multilink $\overrightarrow{m}$ with probability proportional to $\textbf{z}$. In Eq.~(\ref{Xz}),
\begin{equation}\label{kj}
\begin{aligned}
&k_j=\sum_{i=1}^NA_{ij}^{\overrightarrow{m}^{ij}}z^{\overrightarrow{m}^{ij}}+\delta_{0,\sum_{i=1}^NA_{ij}^{\overrightarrow{m}^{ij}}z^{\overrightarrow{m}^{ij}}},\\
&\beta=\frac{1}{N}\sum_{i=1}^N[(1-\widetilde{\alpha})(1-\delta_{0,\sum_{i=1}^NA_{ij}^{\overrightarrow{m}^{ij}}z^{\overrightarrow{m}^{ij}}})+\delta_{0,\sum_{i=1}^NA_{ij}^{\overrightarrow{m}^{ij}}z^{\overrightarrow{m}^{ij}}}]X_j,\\
&v_i=\theta(\sum_{i=1}^NA_{ij}^{\overrightarrow{m}^{ij}}z^{\overrightarrow{m}^{ij}}+\sum_{i=1}^NA_{ji}^{\overrightarrow{m}^{ji}}z^{\overrightarrow{m}^{ij}}).
\end{aligned}
\end{equation}
Here $\delta_{x,y}$ is the Kronecker delta and $\theta(x)$ is the Heaviside step function. The Functional Multiplex PageRank centrality $X_i(\textbf{z})$ of node $i$ depends on the values of $\textbf{z}$. In our calculations for the two layer network, we use a $z^{1,0}=z^{0,1}=z^{1,1}=1$ to represent an equal importance of each type of the multilinks.
\end{definition}
The second multiplex centrality used in this manuscript is the Global Heterogenous Eigenvector-like centrality (GHEC) proposed in ref.~\cite{sola2016}. This centrality takes into account the contribution of neighbors from all layers and the mutual influence between layers. Consider an $M*M$ influence matrix $W$ where the non-negative elements $w_{\alpha\beta}$ represents the influence of the layer $\beta$ on layer $\alpha$, which is
\begin{equation}
        W = \begin{pmatrix}
            \begin{array}{c|c|c}
                    w_{11}  & \cdots & w_{1M}\\ \hline
                    \vdots & \ddots  & \vdots\\ \hline
                    w_{M1}  & \cdots & w_{M,M}\\
            \end{array}
           \end{pmatrix},
\end{equation}
and a $N*NM$ matrix $A=(A_1|A_2|...|A_M)$.
Then $A^{\otimes}$ is the Khatri-Rao product of the matrices $W$ and $A$, which is
\begin{equation}
        A^{\otimes}= \begin{pmatrix}
                       \begin{array}{c|c|c|c}
                    w_{11}A_1 & w_{12}A_2 & \cdots & w_{1M}A_M\\ \hline
                    w_{21}A_1 & w_{22}A_2 & \cdots & w_{2M}A_M\\ \hline
                    \vdots & \vdots & \ddots  & \vdots\\ \hline
                    w_{M1}A_1 & w_{M2}A_2 & \cdots & w_{MM}A_M\\
                    \end{array}
           \end{pmatrix}.
\end{equation}
The Global Heterogenous Eigenvector-like centrality of the multiplex network $\overrightarrow{G}$ is the positive and normalized eigenvector $c^{\otimes}\in R^{NM}$ of the matrix $A^{\otimes}$. If one introduces the notation
\begin{equation}
        c^{\otimes} = \begin{pmatrix}
                         c_1^{\otimes}\\ \hline
                         c_2^{\otimes}\\ \hline
                         \vdots\\ \hline
                         c_M^{\otimes}\\
                    \end{pmatrix}
\end{equation}
with vectors $c_1^{\otimes},...,c_m^{\otimes}\in R^N$, then the Global Heterogenous Eigenvector-like centrality matrix of $\overrightarrow{G}$ is given by
$C^{\otimes}=(c_1^{\otimes}|c_2^{\otimes}|...|c_M^{\otimes})\in R^{N*M}$. The information contained in the vectorial-type centrality should be aggregated to associate a number to each node, which is $C=\sum_{j=1}^M c_j^{\otimes}$, where $C\in R^{N*1}$ and the ith-row $C_i$ is the Global Heterogeneous Eigenvector-like centrality of node $i$ in the multiplex network.
\subsection{Evaluation methods}
To evaluate the performance of measures in identifying super-spreaders, we use the imprecision function~\cite{kitsak2010} and the Kendall's tau correlation coefficient~\cite{kendall1938}. The imprecision function quantifies how close to the optimal spreading is the average spreading of $pN$ nodes with the highest centrality.
\begin{definition}\rm
Imprecision function. The imprecision function is defined as
\begin{equation}\label{imprecision}
\varepsilon(p)=1-\frac{M(p)}{M_{eff}(p)},
\end{equation}
where $p$ is the fraction of nodes considered ($p\in[0,1]$). $M(p)$ is the average spreading influence of $pN$ nodes with the highest centrality, and $M_{eff}(p)$ is the average spreading influence of $pN$ nodes with the highest spreading influence. A smaller $\varepsilon$ value indicates a more accurate centrality to identify super-spreaders.
\end{definition}
\begin{definition}\rm
Kendall's tau correlation coefficient. The Kendall's tau correlation coefficient is used to measure correlation between two rankings which is defined as
\begin{equation}
\tau(r_1,r_2)=\frac{K(r_1, r_2)-K^{'}(r_1,r_2)}{n(n-1)/2},
\end{equation}
where $K(r_1, r_2)$ is the number of node pairs that appear in the concordant ordering in the ranking lists $r_1$ and $r_2$, and $K^{'}(r_1, r_2)$ is the number of node pairs that appear in the reverse ordering in $r_1$ and $r_2$. $n$ is the number of nodes in each ranking list. In our applications, nodes are ranked by centrality measure in ranking list $r_1$ and nodes in ranking list $r_2$ are ranked by spreading influence, which are obtained by computer simulations. A large $\tau$ indicates that the centrality measure and the node spreading influence are highly correlated.
\end{definition}

\section{The proposed coupling-sensitive centrality}
In this part, we first show the two-area phenomena of the centralities to predict the spreading influence of nodes in the contact layer. Then we propose the coupling-sensitive centrality (CS) in multiplex network to predict the disease-spreading influence of nodes in the coevolving dynamics.
\subsection{The two-area phenomena of centrality in predicting disease-spreading influence}
\begin{figure}
\begin{center}
\includegraphics[width=13.5cm]{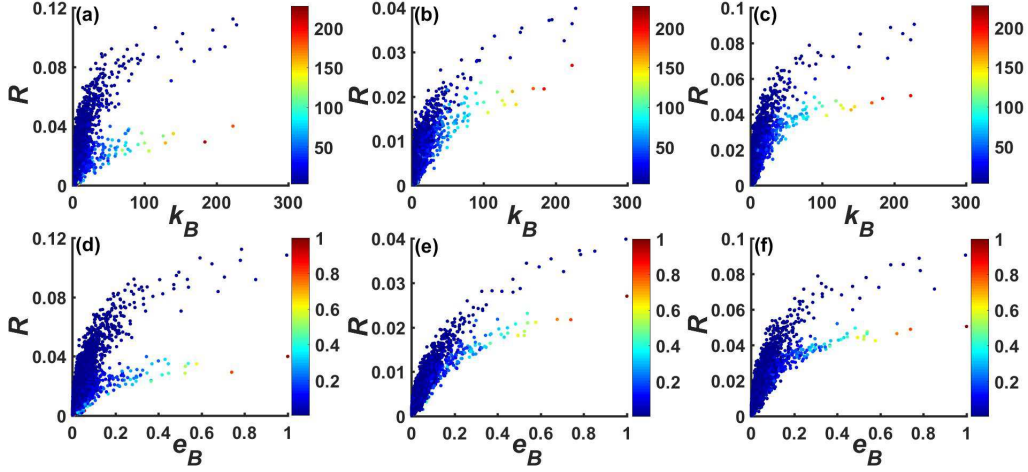}
\caption{The scatter plots of nodes with their disease-spreading influence $R$ and degree centrality $k_B$ (a)-(c) or eigenvector centrality $e_B$ (d)-(f) in the contact layer B. The color of plots represents centralities of nodes in layer A. It can be seen that the spreading influence of nodes with higher centrality in layer A is obviously suppressed. The parameters are set as: (a) and (d) degree correlation $m_s=0.3$, relative spreading rate $\gamma_{AB}^{\lambda}=2.0$, informing rate $\lambda_{BA}=0.1$, and immunization rate $\lambda_{AB}=1.0$; (b) and (e) $m_s=0.5$, $\gamma_{AB}^{\lambda}=2.0$, $\lambda_{BA}=1.0$, $\lambda_{AB}=1.0$; (c) and (f) $m_s=0.5$, $\gamma_{AB}^{\lambda}=3.0$, $\lambda_{BA}=0.1$, $\lambda_{AB}=0.5$. We take these parameter values for example to reflect different dynamical and structural coupling strengths and discuss their impacts later in detail.}
\label{figure2}
\end{center}
\end{figure}

The centrality measures are heuristic ways to predict the spreading influence of nodes in the network. The idea is that the more central a nodes is, the more influential it is in spreading, where the centrality is positively related with the node spreading influence~\cite{jalili2017}. Figure~\ref{figure2} displays the scatter plots of nodes with their disease-spreading influence $R$ and degree centrality $k_B$ or eigenvector centrality $e_B$ on the contact layer B under different parameters. We use three groups of parameter values for the relative spreading rate $\gamma_{AB}^{\lambda}$, the inter-layer dynamic coupling strengths $\lambda_{AB}$ and $\lambda_{BA}$ and the inter-layer degree correlation $m_s$, where the dynamical coupling strength and structural coupling strength vary.

It can be seen that with the increase of $k_B$ and $e_B$, the spreading influence $R$ increases in general. But there are some nodes where their spreading influences are lower than that of others with the same centrality of $k_B$ or $e_B$, and a two-area phenomena appears. The suppressed spreading influence of these nodes is due to the coevolving dynamics between layers. Let's consider two nodes $i_B$ and $j_B$ with the same centrality in layer B and their counterparts in layer A are $i_A$ and $j_A$ respectively. If $i_A$ has a larger centrality than $j_A$, then the spreading of information originating from $i_A$ is supposed to be wider than that originating from $j_A$, making more nodes in layer B be vaccinated. The disease-spreading influence of node $i_B$ in layer B is thus smaller than that of $j_B$. It can be seen from Fig.~\ref{figure2} that for nodes with relatively high centrality in layer A (nodes with color approaching red), their disease-spreading influences are obviously lower than that of nodes with lower centrality in layer A under different parameters. This phenomena implies that the centralities on the contact layer are not adequate to identify the super-spreaders in the coevolving dynamics on the multiplex networks.
\subsection{The coupling-sensitive centrality on multiplex network}
We take into account the dynamical and structural couplings between layers and propose a coupling-sensitive (CS) centrality to identify super-spreaders on multiplex networks.
\begin{definition}\rm
Coupling-sensitive centrality. The coupling-sensitive centrality of node $i$ is defined as
\begin{equation}\label{cs}
CS_i^{\theta}=\theta_i^B-\theta_i^A*\lambda_A*\lambda_{AB}+\theta_i^B*\lambda_B*\lambda_{BA},
\end{equation}
where $\theta_i^B$ is the benchmark centrality of node $i$ in layer B and $\theta_i^A$ is the centrality of its counterpart node in layer A. $\theta$ can be any centrality measure such as degree, eigenvector centrality and k-shell index. $\lambda_A$ and $\lambda_B$ are the transmission rates in layer A and B respectively. $\lambda_{AB}$ is the immunization rate and $\lambda_{BA}$ is the informing rate.\end{definition}

In the definition of CS centrality, the first term on the right side of Eq.~(\ref{cs}) is the node centrality in the contact layer, which is taken as the baseline. The second term represents the impact of information spreading on disease spreading, where the suppression effect of node $i$ depends on the transmission rate $\lambda_A$ in layer A and the inter-layer coupling strength $\lambda_{AB}$ from layer A to layer B. $\theta_i^A$ and $\theta_i^B$ are the centralities of node $i$ in layer A and B respectively, representing the spreading abilities of node $i$ in layer A and B respectively. Existing researches have indicated that the degree centrality is positively correlated with many other centralities, such as the eigenvector centrality and k-shell centrality, so $\theta_i^A$ and $\theta_i^B$ in Eq.~(\ref{cs}) reflect the inter-layer degree correlation. As the information-spreading suppresses the disease-spreading, the spreading influence of node $i$ in layer A has a negative impact on the baseline centrality $\theta_i^B$ in predicting the disease-spreading, thus the second term has a minus sign. On the other hand, the disease spreading in layer B has an impact on layer A through inter-layer coupling, which depends on the transmission rate $\lambda_B$ and the inter-layer coupling strength $\lambda_{BA}$. The more nodes informed by infected nodes in layer B, the less impact of the spreading origin in layer A on information spreading. Thus we add the third term representing the reduced impact of node centrality in layer A. The algorithm to calculate the coupling-sensitive centrality is shown in Algorithm~\ref{alg}.
\begin{algorithm}
\caption{Algorithm to obtain coupling-sensitive centrality}
\label{alg}
\begin{algorithmic}[1]
\REQUIRE ~~
A multiplex network $\overrightarrow{G}=(G_1, G_2)$, where $G_1=(V, E_1)$ and $G_2=(V, E_2)$     //The multiplex network consists of a set of nodes V and two types of edges $E_1$ and $E_2$ which are layer A and layer B respectively.\\
\ENSURE ~~
The coupling-sensitive centrality of each node $CS_i^{\theta}$;
\STATE Set the values of parameters $\lambda_A$, $\lambda_B$, $\lambda_{AB}$ and $\lambda_{BA}$.
\FOR{$i=1$ to $|V|$}
\STATE Calculate the centrality $\theta_i^B$ of node $i$ in layer B;
\ENDFOR
\FOR{$i=1$ to $|V|$}
\STATE Calculate the centrality $\theta_i^A$ of node $i$ in layer A;
\ENDFOR
\FOR{$i=1$ to $|V|$}
\STATE Calculate the centrality $CS_i^{\theta}$ of node $i$ in the multiplex network using Eq.~(\ref{cs})
\STATE $CS_i^{\theta}=\theta_i^B-\theta_i^A*\lambda_A*\lambda_{AB}+\theta_i^B*\lambda_B*\lambda_{BA}$
\ENDFOR
\end{algorithmic}
\end{algorithm}
\section{Experimental results}
We evaluate the performance of the CS centrality and compare it with that of the benchmark centralities in synthesized networks and real-world networks by large simulations. Results show that the proposed CS centrality outperforms the corresponding benchmark single-layer centrality and two typical centralities defined on multiplex networks in identifying super-spreaders in information-disease coevolving process.
\subsection{Performance comparison of coupling-sensitive centrality and single-layer centrality}
\begin{figure}
\begin{center}
\includegraphics[width=13.5cm]{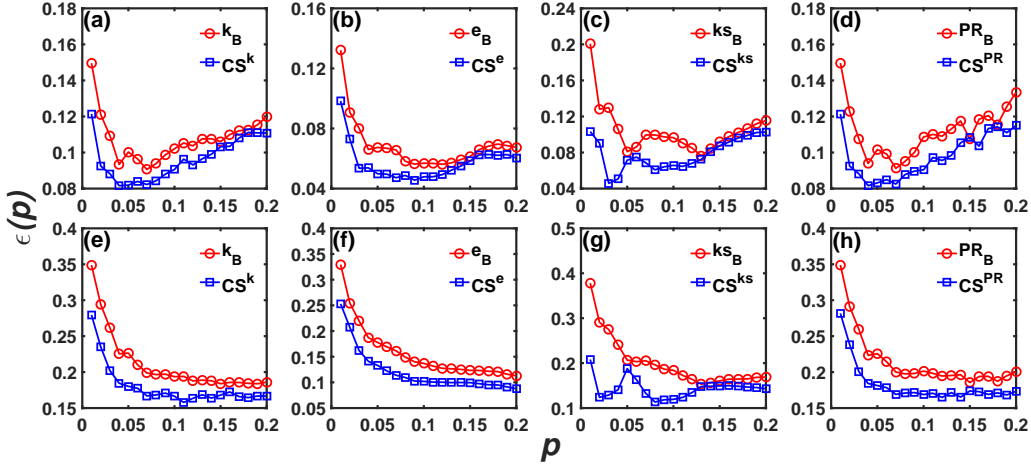}
\caption{Comparison of the imprecisions of benchmark centralities and coupling-sensitive centralities as a function of $p$. $p$ is the proportion of nodes considered. (a)-(d) The parameters are set as $m_s=0.5$, $\gamma_{AB}^{\lambda}=2.0$, $\lambda_{BA}=0.3$, $\lambda_{AB}=0.7$. (e)-(h) The parameters are set as $m_s=0.5$, $\gamma_{AB}^{\lambda}=2.0$, $\lambda_{BA}=0.1$, $\lambda_{AB}=1.0$. The smaller imprecision, the more accurate the measure is in predicting the spreading influence.}
\label{figure3}
\end{center}
\end{figure}

We first compare the performance of CS centrality with that of the single-layer centrality. We use degree centrality $k_B$, eigenvector centrality $e_B$, k-shell index $ks_B$ and PageRank centrality $PR_B$ in layer B as the benchmark centralities respectively and calculate the corresponding CS centralities, which are $CS^{k}$, $CS^{e}$, $CS^{ks}$ and $CS^{PR}$. The imprecisions of these centralities are shown in Fig.~\ref{figure3}.
It can be seen from Figs.~\ref{figure3} (a)-(d) that the CS centralities are more accurate to identify super-spreaders in disease spreading than their corresponding benchmark centralities in layer B. Similar results are obtained under a different set of parameters as shown in Figs.~\ref{figure3} (e)-(f). We use the relatively large $\lambda_{AB}$ and small $\lambda_{BA}$ to emphasize on the cases when the spreading influences of nodes in epidemic are strongly impacted by the information spreading.
\subsection{Performance comparison of coupling-sensitive centrality and multiplex centrality}
\begin{figure}
\begin{center}
\includegraphics[width=13.5cm]{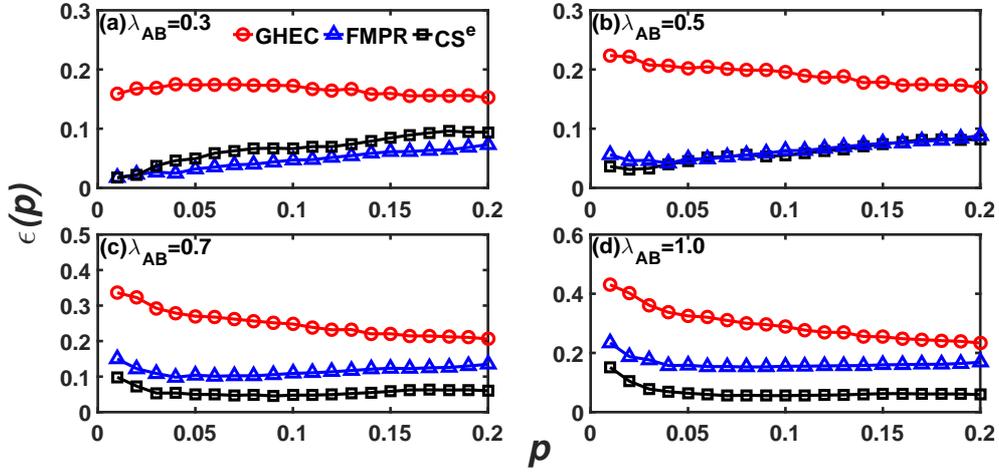}
\caption{Imprecisions as a function of $p$ for the multiplex eigenvector centrality GHEC, multiplex PageRank FMPR and coupling-sensitive centrality $CS^{e}$ based on eigenvector centrality in identifying super-spreaders when the immunization rate $\lambda_{AB}$ varies.}
\label{figure4}
\end{center}
\end{figure}

\begin{figure}
\begin{center}
\includegraphics[width=13.5cm]{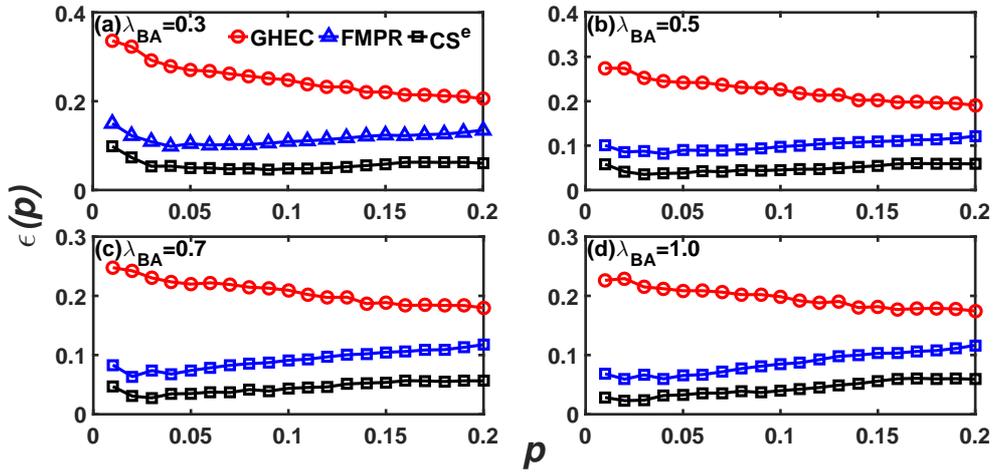}
\caption{Imprecisions as a function of $p$ for the multiplex eigenvector centrality GHEC, multiplex PageRank FMPR and coupling-sensitive centrality $CS^{e}$ in identifying super-spreaders when the informing rate $\lambda_{BA}$ varies.}
\label{figure5}
\end{center}
\end{figure}

\begin{figure}
\begin{center}
\includegraphics[width=13.5cm]{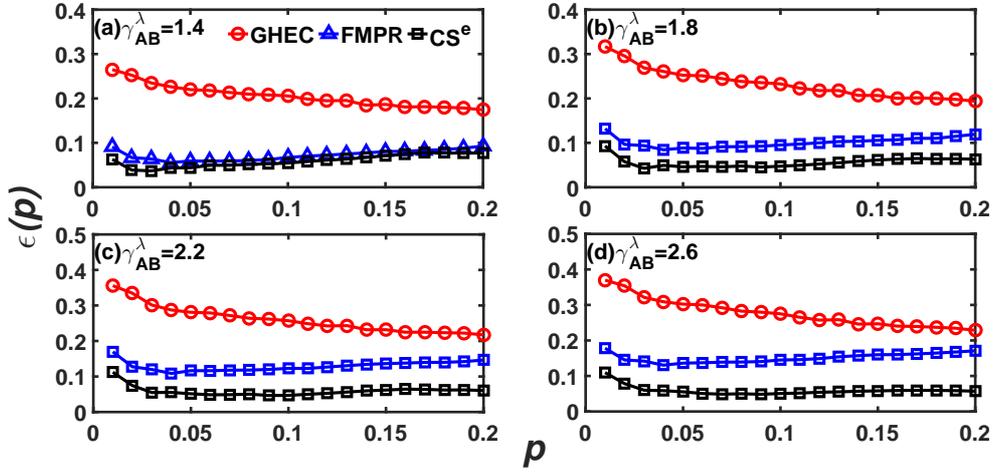}
\caption{Imprecisions as a function of $p$ for the multiplex eigenvector centrality GHEC, multiplex PageRank FMPR and coupling-sensitive centrality $CS^{e}$ in identifying super-spreaders when the relative spreading rate $\gamma_{AB}^{\lambda}$ varies.}
\label{figure6}
\end{center}
\end{figure}

\begin{figure}
\begin{center}
\includegraphics[width=13.5cm]{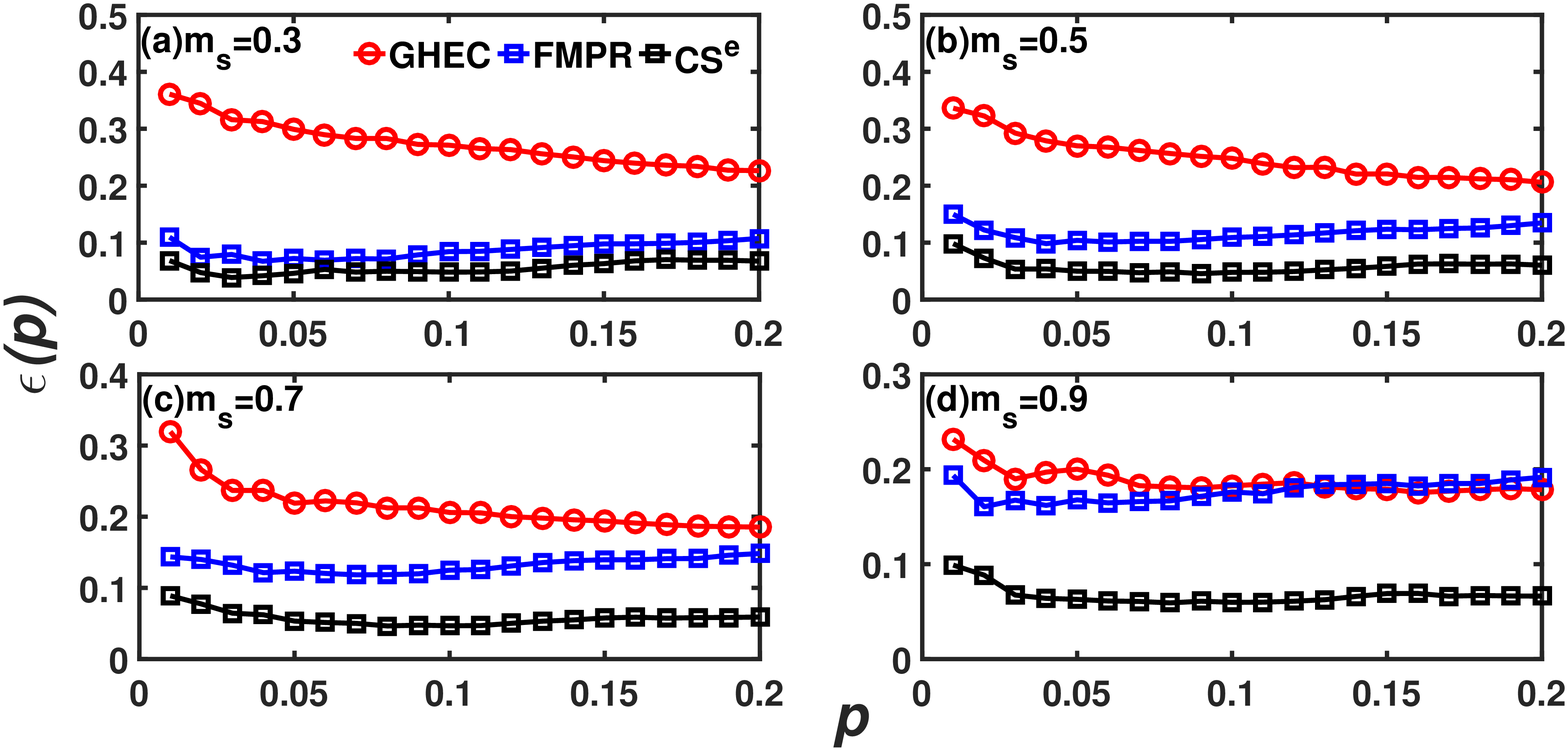}
\caption{Imprecisions as a function of $p$ for the multiplex eigenvector centrality GHEC, multiplex PageRank FMPR and coupling-sensitive centrality $CS^{e}$ in identifying super-spreaders when the inter-layer degree correlation $m_s$ varies.}
\label{figure7}
\end{center}
\end{figure}

Next we explore and compare the performance of CS centrality with two centralities defined on multiplex networks. We use the eigenvector centrality as the benchmark centrality to calculate the coupling-sensitive centrality $CS^{e}$, and compare with the Functional Multiplex PageRank~\cite{iaco2016} and the Global Heterogeneous Eigenvector-like centrality~\cite{sola2016}. To validate the robustness of the proposed CS centrality, we vary the structural and dynamical coupling strengths, which are represented by the immunization rate $\lambda_{AB}$, the informing rate $\lambda_{BA}$, the relative spreading rate $\gamma_{AB}^{\lambda}$ and the inter-layer degree correlation $m_s$. Results indicate that the CS centrality is in general robust under all considered parameters, as shown in Fig.~\ref{figure4}-Fig.\ref{figure7}.

Figure \ref{figure4} displays the imprecisions of centralities when the immunization rate $\lambda_{AB}$ varies. When $\lambda_{AB}\leq0.5$, the imprecision of $CS^e$ is a little higher than or close to that of the multiplex centrality FMPR. When $\lambda_{AB}$ increases, the $CS^e$ outperforms the other two centralities. This is because when the immunization rate $\lambda_{AB}$ is relatively small, according to the definition of CS centrality, the second term is relatively small, and the CS centrality is more close to the centrality of node in layer B. When $\lambda_{AB}$ becomes large, the CS centrality is more impacted by the centrality $\theta_i^A$ of node $i$ in layer A. In this case the CS centrality can reflect the contributions of both layers, thus is more effective.
While the informing rate $\lambda_{BA}$, the relative spreading rate $\gamma_{AB}^{\lambda}$ and degree correlation $m_s$ vary respectively, as shown in Fig.~\ref{figure5}-Fig.\ref{figure7}, the imprecision of $CS^e$ is the smallest, which indicates that the coupling-sensitive centrality is the best measure to identifying the super-spreaders in the coevolving dynamics.

\begin{figure}
\begin{center}
\includegraphics[width=13.5cm]{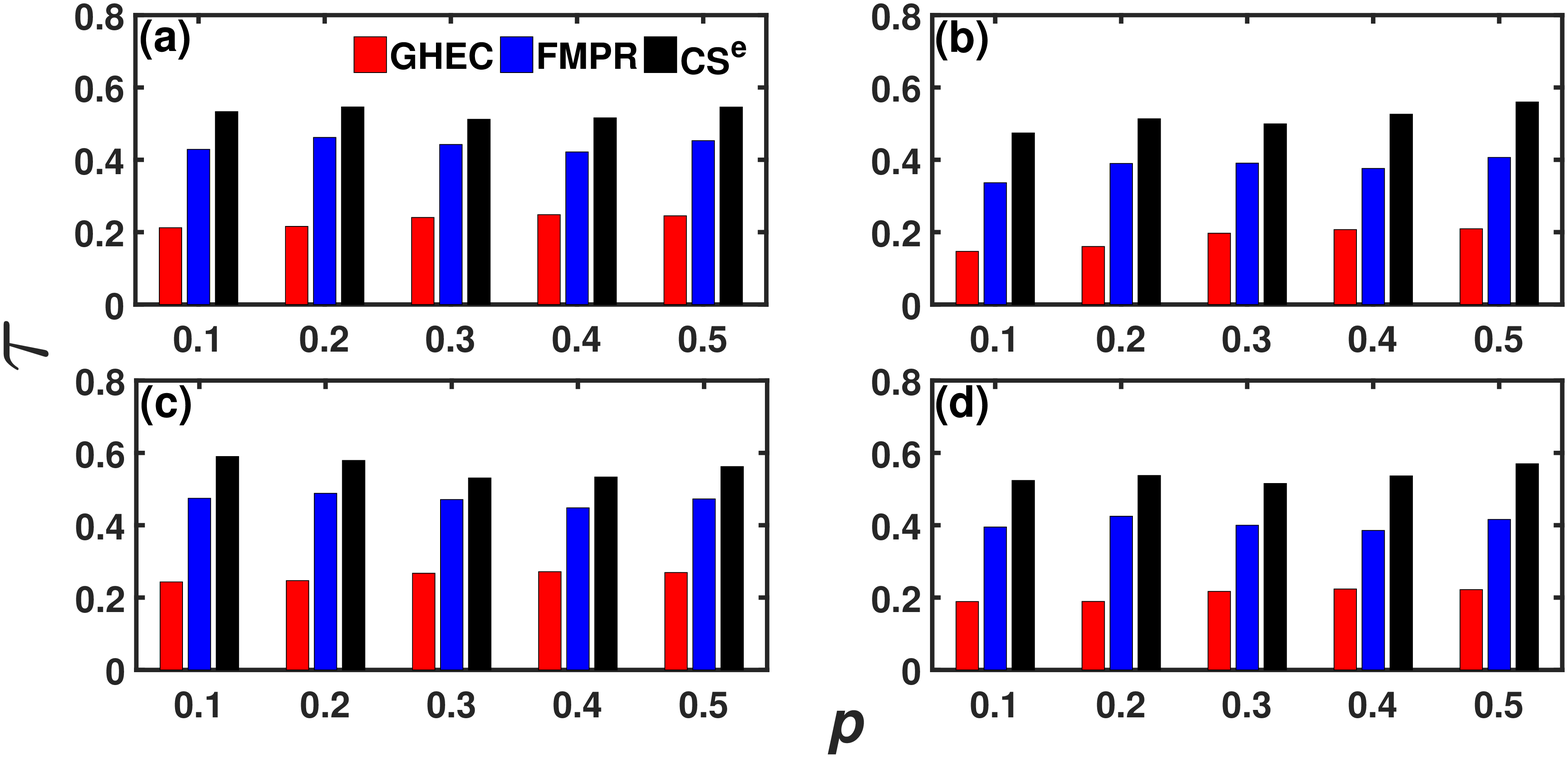}
\caption{Kendall's tau correlation of the centrality measure and the spreading influence of nodes under different parameters. The centralities used are the multiplex eigenvector centrality GHEC, multiplex PageRank FMPR and coupling-sensitive centrality $CS^{e}$. $p$ is the proportion of nodes considered. (a)$m_s=0.5$, $\lambda_{AB}=1.0$, $\lambda_{BA}=0.1$, $\gamma_{AB}^{\lambda}=2.0$. (b) $m_s=0.5$, $\lambda_{AB}=0.6$, $\lambda_{BA}=0.1$, $\gamma_{AB}^{\lambda}=2.0$. (c) $m_s=0.5$, $\lambda_{AB}=1.0$, $\lambda_{BA}=1.0$, $\gamma_{AB}^{\lambda}=2.0$. (d) $m_s=0.5$, $\lambda_{AB}=1.0$, $\lambda_{BA}=0.1$, $\gamma_{AB}^{\lambda}=1.6$.}
\label{figure8}
\end{center}
\end{figure}

Furthermore, we calculate the Kendall's tau correlation coefficient of the proposed CS centrality with the spreading influence of nodes, and compare with that of the multiplex centralities. It be can be seen from Fig.~\ref{figure8} that the CS centrality has the largest correlation with the spreading influences of the influential spreaders.
\subsection{Application in real-world networks}
Many real-world networks can be represented as multilayer networks, such as the biological networks, the social networks and the cooperation networks. In this part, we apply the coupling-sensitive centrality in eight real-world multilayer networks and compare its performance with other two multiplex centralities. For the multilayer networks with more than two layers, we choose two layers and use the mutually connected giant component in which the nodes are connected to the largest connected component in both layers for study. The real-world networks studied are: (1) SacchPomb (gene and protein interaction network. Layer A is a synthetic genetic interaction network and layer B is a physical association network); (2) Drosophila (protein-protein network with different nature of interactions as layers. Layer A is suppressive genetic interaction and layer B is additive genetic interaction); (3) IPv4\_IPv6 (Internet topologies of Autonomous System (AS). Layer A is IPv4 AS network and layer B is IPv6 AS network); (4) Human brain (structural and functional networks where nodes are brain regions. Layer A is the structural network and layer B is the functional network ); (5) Physicians (different types of relationships among physicians in four US towns. Layer A is a discussion network and layer B is a friendship network); (6) arXiv (co-authorship of scientists in different research categories. Layer A is a co-authorship network in data analysis, statistics and probability and layer B is a co-authorship network in physics and society); (7) Air\_train (airports network and train stations network in India. Layer A is the train network and layer B is the airport network); (8) Pardus (friendship relationship and message communication relationship between individuals in an online game virtual society. Layer A is the friendship network and layer B is the message communication network). Data for the first seven networks are collected from ref.~\cite{klein2016} and there are more detailed descriptions of the data and their origins, and data for the last network are collected from the massive multiplayer online game $'Pardus'$(http://www.pardus.at)~\cite{szell2010}. The characteristics of the studied networks are listed in Table~\ref{basiccharacteristic}.
\begin{table}[htbp]
\begin{center}
\footnotesize
\caption{\label{basiccharacteristic}Characteristics of the real-world multiplex networks studied in this work. These characteristics include the number of nodes $N$, the number of edges $E_A$ in layer A and $E_B$ in layer B, the inter-layer degree correlation of two layers $m_s$, the infection rate $\lambda_B$ used in the SIR spreading in layer B, and the type of networks.}
\begin{tabular}{cccccccccccc}
\hline
Network & \textbf{$N$} & \textbf{$E_A$} & \textbf{$E_B$} & \textbf{$m_s$} & \textbf{$\lambda_B$} & Type\\
\hline
SacchPomb  &426 &2236 &1678 &0.27 &0.28 &Biological\\
Drosophila &449 &2656 &2172 &0.65 &0.21 &Biological\\
IPv4\_IPv6  &4710 &48026 &25366 &0.55 &0.018 &Technological\\
Human brain &74 &426 &396 &0.24 &0.517 &Biological\\
Physicians  &106 &460 &362 &0.47 &0.818 &Social\\
arXiv  &2252 &15926 &14570 &0.94 &0.24 &Collaboration\\
Air\_train  &66 &634 &354 &0.44 &0.185 &Technological\\
Pardus  &2501 &16119 &20860 &0.65 &0.052 &Social\\
\hline
\end{tabular}
\end{center}
\end{table}

\begin{figure}
\begin{center}
\includegraphics[width=13.5cm]{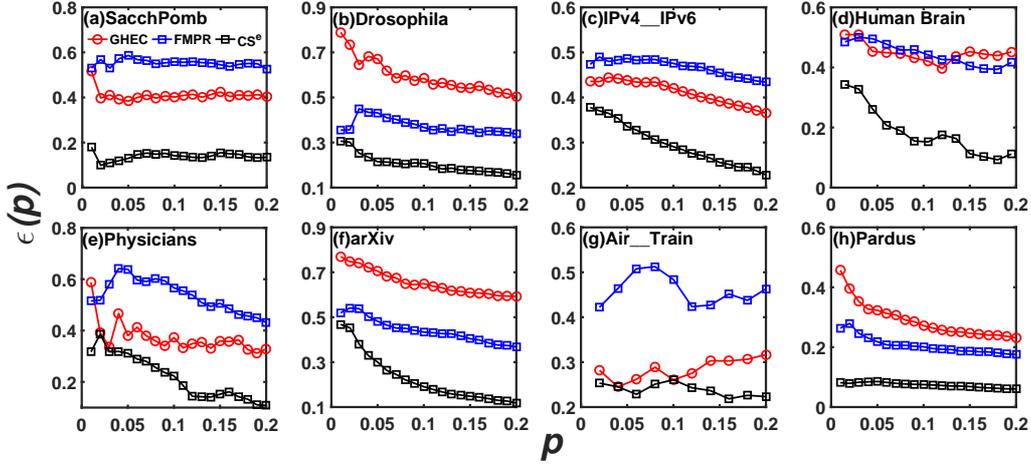}
\caption{Imprecisions of the centralities on real-world multiplex networks as a function of $p$. The centralities used are the multiplex eigenvector centrality GHEC, multiplex PageRank FMPR and coupling-sensitive centrality $CS^{e}$. The parameters are set as $\lambda_{AB}=0.7$, $\lambda_{BA}=0.3$, $\gamma_{AB}^{\lambda}=2.0$. $\lambda_B$ for each network is three times of the epidemic threshold of layer B.}
\label{figure9}
\end{center}
\end{figure}

It can be seen from Fig.~\ref{figure9} that in all studied networks, the coupling-sensitive centrality outperforms the other two multiplex centralities. This implies that considering the structural and dynamical couplings between layers is significantly meaningful to identify the most important nodes in the multilayer networks.

\section{Conclusion and discussion}
Identifying the most influential spreaders is an important step to make use of available resource and control the spreading process. In this paper, we work on identifying super-spreaders in the information-epidemic coupled spreading dynamics on multiplex networks. We find that the centralities on the contact layer are no longer effective to identify the most influential disease-spreaders, which is due to the suppression effects from the information layer. By considering the centralities of nodes in both layers and the structural and dynamical couplings between layers, we propose a new measure called coupling-sensitive centrality on multiplex networks. Simulation results on synthesized networks and real-world networks indicate that the CS centrality is not only much more effective than the centralities on the single contact layer, but also more effective than two typical multiplex centralities without considering the dynamics.

The real-world multiplex networks studied in our simulations are not necessarily the networks where information and disease spread. But we can consider that the dynamic couplings in our measure have their specific meanings in different contexts of the various real-world networks. For example, the immunization rate $\lambda_{AB}$ may represent the proportion of autonomous systems (ASs) routing IPv4 packets that can also route IPv6 packets in the IPv4\_IPv6 multiplex network, and the informing rate $\lambda_{BA}$ may represent the proportion of ASs routing IPv6 packets that can also route IPv4 packets. While in the air\_train network, the immunization rate $\lambda_{AB}$ may represent the possibility that the travellers planning to travel by airplane switch to travel by train, and the informing rate $\lambda_{BA}$ may represent the possibility of travellers changing the transportation mode from train to airplane. The effectiveness of the coupling-sensitive centrality implies that when identifying the most important nodes in the multilayer systems, considering their structural and dynamical couplings are very necessary. Our work gives a possible way to synthesize the inter-layer couplings into one measure which may find its applications in various multilayer systems.

The computation of CS centrality contains the spreading rates of two layers, which are the immunization rate and the informing rate. Their values vary depending on the specific real-world scenarios of information spreading and epidemic spreading. For example the basic reproduction number $R_0$, which is the average number of secondary infections caused by a primary case, is used to estimate the spreading parameters, where $R_0=\lambda/\mu$~\cite{anderson1992}. Our work demonstrates that in a reasonable parameter range, for example the information spreading rate is larger than the epidemic spreading rate corresponding to $\gamma_{AB}^\lambda>1$, the CS centrality is effective and robust under different values of the parameters $\lambda_{AB}$ and $\lambda_{BA}$. In addition, in the definition of the coupling-sensitive centrality, there are two items $\theta_i^A*\lambda_A*\lambda_{AB}$ and $\theta_i^B*\lambda_B*\lambda_{BA}$ that represent the mutual influence of two layers. If we add two weighting factors $\alpha$ and $\beta$ before each item, adjusting them may come to an optimal expression of the coupling-sensitive centrality.

We work on the two-layer multiplex network where the information spreading and epidemic spreading coevolves on each layer. For networks with more than two layers the identification of critical nodes will be more complex due to the coupling and mutual influences between layers. The study of more general index in multilayer networks requires further explore in the future.
\section*{Acknowledgments}
This work is supported by the National Natural Science Foundation of China (Grant Nos. 61802321, 11975099), the Sichuan Science and Technology Program (No. 2020YJ0125), the Natural Science Foundation of Shanghai (Grant No.~18ZR1412200) and the Southwest Petroleum University Innovation Base (No.642).

\end{document}